\documentclass[conference]{IEEEtran}

\usepackage{verbatim}
\usepackage{epsfig,bbm}
\usepackage{CJK}
\usepackage{indentfirst}
\usepackage{multirow}
\usepackage{epstopdf}
\usepackage{graphicx}
\usepackage{footmisc}
\graphicspath{{./figures/}}
\usepackage{amsfonts}
\usepackage{mathrsfs}
\usepackage{setspace}
\usepackage{amsmath}
\usepackage{algorithm,algorithmic,amsbsy,amsmath,amssymb,epsfig,bbm,mathrsfs, bbm} 
\usepackage{amsthm}
\usepackage{verbatim}
\usepackage[subfigure]{graphfig}

\begin{document}

\title{Hierarchical Cooperation for Operator-Controlled Device-to-Device Communications: A Layered Coalitional Game Approach \vspace{-9mm} }

\author{\\  Xiao Lu, Ping Wang, Dusit Niyato\\
   ~School of Computer Engineering, Nanyang Technological University, Singapore \\
  Email: \{Luxiao, Wangping, Dniyato\}@ntu.edu.sg \vspace{-6mm}
  }

\markboth{}{Shell \MakeLowercase{\textit{et al.}}: Bare Demo of
IEEEtran.cls for Journals}

\maketitle

\begin{abstract}
 
Device-to-Device (D2D) communications, which allow
direct communication among mobile devices, have been
proposed as an enabler of local services in 3GPP LTE-Advanced
(LTE-A) cellular networks. This work investigates a hierarchical LTE-A network framework consisting of multiple D2D operators at the upper layer and a group of devices at the lower layer. We propose a cooperative model that allows the operators to improve their utility in terms of revenue by sharing their devices, and the devices to improve their payoff in terms of end-to-end throughput by collaboratively performing multi-path routing. To help understanding the interaction among operators and devices, we present a game-theoretic framework to model the cooperation
behavior, and further, we propose a layered coalitional game
(LCG) to address the decision making problems among them.
Specifically, the cooperation of operators is modeled as an
overlapping coalition formation game (CFG) in a partition form, in which operators should form a stable coalitional structure. Moreover, the cooperation of devices is modeled as a coalitional graphical game (CGG), in which devices establish links among each other to form a stable network structure for multi-path routing. We adopt the extended recursive core, and Nash network, as the stability concept for the proposed CFG and CGG, respectively. Numerical results demonstrate that the proposed LCG yields notable gains compared to both the non-cooperative case and a LCG variant and achieves good convergence speed.

\end{abstract}

\emph{Keywords-} D2D communications, LTE-Advanced network, layered coalitional game, coalitional structure formation, multi-path routing, extended recursive core, coalitional graphical game, Nash network. 

\IEEEpeerreviewmaketitle
\section{Introduction}

\newtheorem{definition}{Definition}

Device-to-Device (D2D) communications \cite{K2009,Lei2012L} have emerged as a promising paradigm for 3GPP LTE-Advanced (LTE-A) networks, which provide mobile wireless connectivity, reconfigurable architectures, as well as various wireless applications (e.g., network gaming, social content sharing and vehicular networking) for better user experience.
With D2D communications, nearby devices in a cellular network can communicate with each other directly bypassing the base stations. Conventional D2D communications commonly refer to direct information exchanges among devices in Human-to-Human and Machine-to-Machine communications, without the involvement of wireless operators. However, conventional D2D technologies cannot provide efficient interference management, security control and quality-of-service guarantee \cite{Lei2012L}. Recently, there is a trend towards operator-controlled D2D communications to facilitate profit making for operators as well as better user experience for devices \cite{Lei2012L}.

This paper considers a multi-hop LTE-A network consists of devices deployed by multiple operators. In this network, an efficient approach to improve the end-to-end throughput is to enable cooperative sharing of idle devices among the operators for multi-path routing \cite{Lu2011}. 
The cooperation can increase throughput for the devices because a cooperative relay may substantially lead to improved network capacity. Accordingly, a larger amount of user traffic demand can be supported, which will lead to higher aggregated revenue for operators. In this cooperation, each operator needs to decide on which operators to cooperate with to maximize profit and, given the cooperation behavior of operators. Then, the devices from cooperative operators need to make decision on which devices to cooperate with to maximize throughput.
We call the formation of this interrelated operator cooperation and device cooperation as a hierarchical cooperation problem, which is the main focus of this paper. 
The hierarchical cooperation gives rise to two major concerns. Firstly, what is the stable coalitional structure desirable for all operators so that none of operators is willing to leave the coalition? Secondly, what is the stable network structure for cooperative devices to perform multi-path routing? This paper addresses these two concerns by formulating a layered game framework to model the LTE-A network with operators and devices being the players in the upper and lower layer, respectively. Previous work has also considered game-theoretic framework with hierarchies/layers, e.g., \cite{Xiao2012} in the cognitive radio networks, \cite{Kang2012} in two-tier femtocell networks and \cite{cao2013} in WRNs. However, most of the works considered the competition relationship between different layers, which belongs to the Stackelberg game concept \cite{Fudenberg1993}. In the proposed layered game framework, different layers interact to improve the benefit of each other cooperatively. We adopt the concepts of an extended recursive core and Nash network as the solutions for the proposed games in the upper layer and lower layer, respective. To the best of our knowledge, this is the first work to introduce the application of extended recursive core in wireless communications.
 
\vspace{-5mm}
\section{Network Model and Problem Formulation}

\begin{figure}
\centering
\includegraphics[width=0.3\textwidth]{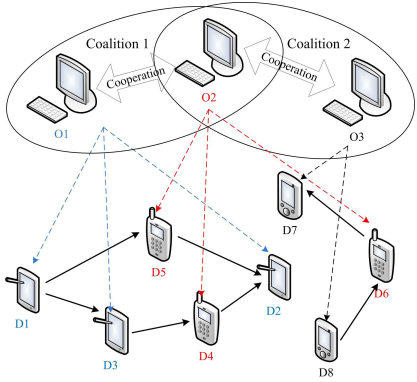}
\caption{System model of Operator controlled D2D Communications} \label{model}
\end{figure}

We consider an LTE-A network consisting of a number of devices belonging to multiple operators. We denote the set of operators as $\mathcal{H}=\{ 1, 2,...,H\}$, and the set of devices of operator $h \in \mathcal{H}$ as $\mathcal{M}^{(h)}=\{ 1, 2,...,M_{h}\}$. The operators are willing to form overlapping coalitions to maximize their individual profits.
An overlapping coalitional structure for a number of operators can be defined in a cover function as the set $\pi_{\mathcal{H}} = \{ \mathcal{S}_{1},\mathcal{S}_{2},\ldots,\mathcal{S}_{z} \}$ which is a collection of non-empty subsets of $\mathcal{H}$ such that $\bigcup^{z}_{k=1} \mathcal{S}_{k} = \mathcal{H}, \forall k, \mathcal{S}_{k} \subseteq \mathcal{H} $ and the sub-coalitions could overlap with each other. $z$ is the total number of coalitions in collection $\pi_{\mathcal{H}}$. Let $\Gamma_{h}$ denote the set of coalitions that operator $h$ belongs to.

For multiple access at every hop, we consider an OFDMA-based transmission\footnote{Other multiple access techniques can be used without loss of generality in the analysis of this paper}. 
In an operator coalition $\mathcal{S} \subseteq \mathcal{H}$, each relay device $i \in \mathcal{M}^{(h)}$ not only needs to support internal flow transmission demand, but also can serve as a relay for other devices from cooperative operators $ h' \in \mathcal{S} \setminus \{h\} $. 
Due to the limited transmission power of each device, multi-hop relaying is adopted to route flow sessions from source devices to destination devices. Since single path routing is too restrictive for satisfying traffic demand, we assume each flow session can be split for multi-path routing if necessary.

Figure \ref{model} illustrates an example for the studied system model and the corresponding notations. In this example, the LTE-A network is composed of $8$ devices deployed by $3$ different operators, i.e., $\mathcal{H}= \{O1, O2, O3\}$, $M^{O1} = \{D1, D2, D3\}$, $M^{O2} = \{D4, D5, D6\}$ and $M^{O3} = \{D7, D8\}$. From Fig. \ref{model}, there is a multi-path flow sourced from $D1$ to $D2$ and a single-path flow sourced from $D8$ to $D7$. From the chosen links, the final coalitional structure of the operators is $\{(O1, O2), (O2, O3)\}$.

Let $\mathcal{L}_{h}=\{1,2,\ldots,l^{(h)}\}$ denote the set of flow sessions from operator $h$. $\mathcal{N}(l^{(h)})$ denote the set of nodes of flow $l^{(h)}$.  The source and destination device of flow $l^{(h)}$ is represented as $s(l^{(h)})$ and $d(l^{(h)})$, respectively.
We denote the link between two devices $i$ and $j$ as $(i,j)$. 


The channel gain on a link $(i,j)$ can be obtained from~\cite{hou08}:
$g_{ij}=\beta \cdot d^{-n}_{ij},
\label{gain}$
where $\beta$ is the antenna related constant, $n$ is the path loss exponent, and $d_{ij}$ is the distance between devices $i$ and $j$.

$f_{(i, j)}(l^{(h)})$ denotes the data rate on link $(i, j)$ attributed to a flow session $l^{(h)} \in \mathcal{L}^{(h)}, h\in \mathcal{S} \subseteq \mathcal{H}$. Since, for D2D communications, a flow session from a source device may traverse multiple relay devices to reach its destination device, we consider the following two cases about a  device.

1) If a  device $i$ is the source or destination of flow session $l^{(h)}$, i.e., $i=s(l^{(h)})$ or $i=d(l^{(h)})$, then
\begin{eqnarray}  \small
\sum_{j \in \mathcal{A}_{i}} f_{(i, j)}(l^{(h)})=r(l^{(h)}) \label{source} \hspace{2mm} \text{or} \sum_{p \in \mathcal{A}_{i}} f_{(p, i)}(l^{(h)})=r(l^{(h)}), \label{destination} 
\end{eqnarray}
where $r(l^{(h)})$ is the aggregated rate of flow session $l^{(h)}$, and $\mathcal{A}_{i}$ denotes the set of devices having direct link with $i$.

2) If a  device $i$ is an intermediate relay device of flow session $l^{(h)}$, i.e., $i \neq s(l^{(h)})$ and $i \neq d(l^{(h)})$, then
\begin{eqnarray} 
\sum^{j \neq s(l^{(h)})}_{j \in \mathcal{A}_{i}} f_{(i, j)}(l^{(h)})=\sum^{p \neq d(l^{(h)})}_{p \in \mathcal{A}_{i}} f_{(p, i)}(l^{(h)}). \label{intermediate}
\end{eqnarray}

Let $c_{ij}$ denote the capacity of link $(i ,j)$. The aggregated data rate on each link $(i,j)$ cannot exceed the link's capacity. Thus we have the following constraint,
\begin{eqnarray} 
\label{rate} 
\sum_{\mathcal{S} \in \Gamma_{h}} \sum_{h \in \mathcal{S}} \sum^{i \neq d(l^{(h)}), j \neq s(l^{(h)})}_{l^{(h)}\in \mathcal{L}^{(h)}}f_{(i, j)}(l^{(h)}) \leq  c_{ij}.
\end{eqnarray}

Let $f^{\star}_{(i, j)}(l^{(h)})$ denote the maximal rate a flow session $l^{(h)}$ that is available on link $(i, j)$, with the constraints in (\ref{source}), (\ref{intermediate}) and (\ref{rate}), and $F^{\star}(l^{(h)})$ the maximal aggregated rate of a flow session $l^{(h)}$. We have $F^{\star}(l^{(h)})= \sum^{i = s(l^{(h)})}_{j \in \mathcal{S}_{i}} f^{\star}_{(i, j)}(l^{(h)})$.
The aggregated rate of flow session $l^{(h)} \in \mathcal{L}^{(h)}$ is constrained by
\begin{eqnarray}
r(l^{(h)})= \left\{
\begin{array}{rcl}
D(l^{(h)})  ,  & &   { D(l^{(h)})     \leq  F^{\star}(l^{(h)}) },    \\
 F^{\star}(l^{(h)}) , & &  \text{Otherwise},
\end{array} \right. \end{eqnarray}
where $D(l^{(h)})$ is the rate demand of flow session $l^{(h)}$.

\section{Layered Coalitional Game for Hierarchical LTE-A networks}

\subsection{Layered Coalitional Game Framework}
We formulate a game-theoretic framework, referred to as the layered coalitional game (LCG), to model the decision-making process of hierarchical cooperation between the operators and devices. Both operators and devices are assumed to be self-interested and rational. The operators aim to maximize their individual utility, while the devices attempt to maximize their end-to-end throughput with the help of relay devices from cooperative operators. In this LCG, the operators need to decide on the coalitional structure and the devices need to make decisions to form a relay network structure, both in distributed ways, with the aim to improve their utilities and payoffs respectively. As both operators and devices only have limited information at their own layer, information exchange between both layers is required. 
The operators need to collect the payoff information from devices, to make the decision of coalitional structure formation. The decision of operators will then be provided to the devices for the purpose of network structure formation. In this regard, there could be multiple interactions between the operator layer and device layer. Recognizing the behavior of operators and devices, we propose to use the overlapping coalition formation game (CFG) to model the behavior of operators and the coalitional graphical game (CGG) to characterize the interaction among devices, which will be introduced in Section III-B and Section III-C, respectively.

Since operators and devices have different objectives and concerns during the cooperation, our next step is to define the objective functions to capture the incentives for operators and devices. 
For a given operator coalition $\mathcal{S}$, we define the payoff of a  device $ i \in \mathcal{M}^{(h)}$ from operator $h \in \mathcal{S}$, which performs flow transmission, as the end-to-end throughput of the flow from device $i$ to device $k$, which is expressed as follows,

\begin{eqnarray} \label{device} \small   
 v(\{i\})= \sum_{\mathcal{S} \in \Gamma_{h}} \sum_{h \in \mathcal{S}} \sum_{l^{(h)} \in \mathcal{L}^{h}}  \sum^{i = n(l^{(h)})}_{j \in \mathcal{A}_{i}}   f_{(i, j)}(l^{(h)})  \\ 
\text{with the constraints in (2), (3), (4) } \nonumber
\end{eqnarray}
where $n(l^{(h)}) \in \mathcal{N}(\mathcal{L}^{(h)})$.

A relay device $j$ aims to help a  device $i$ on transmission to improve the throughput by most. Therefore, to evaluate how much the relay device $j$ can help to improve the throughput of the device $i$, we define the payoff of the relay device $j$ as the difference between the throughput of the device $i$ with the help of device $j$ and that without the help of device $j$, which can be expressed as follows:

\begin{flalign}\label{Rdevice}
 v(\{j\}) &= v^{(j)}(\{i\})-v^{(/j)}(\{i\}) \nonumber  \\
&= \sum_{\mathcal{S} \in \Gamma_{h}} \sum_{h \in \mathcal{S}} \sum_{l^{(h)} \in \mathcal{L}^{h}}  \sum^{i = n(l^{(h)})   }_{j \in \mathcal{A}_{i} \cup \{j\} }   f_{(i, j)}(l^{(h)})  \nonumber \\ 
& - \sum_{\mathcal{S} \in \Gamma_{h}} \sum_{h \in \mathcal{S}} \sum_{l^{(h)} \in \mathcal{L}^{h}} \sum^{i = n(l^{(h)})}_{j^{\prime} \in \mathcal{A}_{i}}  f_{(i, j)}(l^{(h)}) 
\end{flalign}
\begin{eqnarray}
\hspace{10mm}
  \text{with the constraints in (2), (3), (4) } \nonumber
\end{eqnarray}
 
where 
$v^{(j)}(\{i\})$ represents the payoff of device $i$ with the help of device $j$ and  $v^{(/j)}(\{i\})$ represents the payoff of device $i$ without the help of device $k$. 
In the proposed LCG, operators are allowed to form overlapping coalitions to share their idle devices as relays with the aim to improve the aggregated throughput.

Through multi-path routing, each operator aims to improve the aggregated throughput as much as possible. Thus, in return, higher revenue for providing the flow to meet the customer demand will be rewarded for each operator. We define the individual utility of the operator as the profit, i.e., revenue minus the cost of devices in transmitting and forwarding a packet (e.g., due to energy consumption). Let $P_{R}$ denote the rewarded revenue per unit throughput achieved per unit time, and $P^{C}_{i}$ denote the operation cost per device $i$ per unit time. In this case, we assume $P_{R} >> P^{C}_{i}$. 
Then, given a partition $\pi_{\mathcal{H}} $, for an operator $h$ without cooperation, i.e., $\{h\} \in \pi_{\mathcal{H}}$, the utility function $u^{\pi_{\mathcal{H}}}(\{h\})$ can be calculated as (\ref{operator}), 
\begin{eqnarray} \label{operator}
u^{\pi_{\mathcal{H}}}(\{h\})= \sum_{l^{(h)} \in \mathcal{L}^{(h)}} \sum^{i = s(l^{(h)})}_{j \in \mathcal{A}_{i}}   f_{(i, j)}(l^{(h)}) \cdot P_{R}  \nonumber \\ -  
\sum_{ l^{(h)} \in \mathcal{L}^{(h)}}  \sum_{i \in \mathcal{N}(l^{(h)})}  P^{C}_{i}. \\
\text{with the constraints in (2), (3), (4) } \nonumber
\end{eqnarray}
The first term and the second term on the right side of (\ref{operator}) represents the aggregated revenue for operator $h$ and the total device costs, respectively.

While cooperation can lead to profit improvement for operators, it may also incur inherent coordination costs, such as packet overhead. Let $\xi^{(\mathcal{S}_{h})}$ denote the coalition cost incurs to operator $h$ for being coaliton $\mathcal{S}$.
The objective for operator cooperation through cooperative sharing of devices, is to maximize their aggregated profit, i.e., revenue is subtracted by device operation cost and coalition cost. Given the partition $\pi_{\mathcal{H}}$, we define the utility of an operator coalition $\mathcal{S} \in \pi_{\mathcal{H}}$ as the profit of the coalition as follows,
\begin{eqnarray}\footnotesize \label{utility}
u^{\pi_{\mathcal{H}}}(\mathcal{S})= \sum_{h \in \mathcal{S}} \sum_{l^{(h)} \in \mathcal{L}^{(h)}} \sum^{i = s(l^{(h)})}_{j \in \mathcal{A}_{i}}   f_{(i, j)}(l^{(h)}) \cdot P_{R}  \nonumber \\ -  
\sum_{h \in \mathcal{S}} \sum_{ l^{(h)} \in \mathcal{L}^{(h)}}  \sum_{i \in \mathcal{N}(l^{(h)})}  P^{C}_{i}- \sum_{h \in \mathcal{S}} \xi^{(\mathcal{S})}_{h} \\
\text{with the constraints in (2), (3), (4) } \nonumber
\end{eqnarray}


\subsection{Overlapping Coalition Formation Game}

An overlapping CFG is formulated among operators whose interests are to satisfy its internal flows with as less cost as possible. Due to interference, the utility of any operator is affected by not only the behavior of others in the same coalition, but also that of operators from other coalitions. Thus, the considered operator coalitional game is in a partition form since the aggregated utility of a coalition $\mathcal{S} \in \pi_{\mathcal{H}}$ depends on the coalitional structure $\pi_{\mathcal{H}}$ of all the operators $\mathcal{H}$ in the network.
We introduce the framework of an overlapping CFG in partition form with non-transferable utility to model the cooperation among operators. 

\begin{definition}
An overlapping CFG in partition form with non-transferable utility (NTU) is defined by a pair $(\mathcal{H},u)$ where $\mathcal{H}$ is the set of players, and $u$ is a value function that maps every partition $\pi_{\mathcal{H}}$ and every coalition $\mathcal{S} \in \mathcal{H}, \mathcal{S} \in \pi_{\mathcal{H}}$ to a real number that represents the total utility (profit) that players in coalition $\mathcal{S}$ can obtain.    
\end{definition}

The \emph{strategy} of an operator is to form the coalitions to improve its individual utility defined by (\ref{operator}). Note that different from non-overlapping CFG where players have to cooperate with all others in the same coalition and each player only stays in one coalition, in an overlapping CFG, each player is able to join multiple different coalitions.

The solution of the overlapping CFG is the stable overlapping coalitional structure for operators, under which no one will deviate. To this end, we adopt the concept of \emph{extended recursive core} \cite{Agbaglah2009}, referred to as $\gamma^{\dagger}\text{-}core$, as the solution. $\gamma^{\dagger}\text{-}core$ is an extended solution of coalition formation game, which allows coalitions to overlap, accounting for externalities across coalitions. In the proposed game, the externalities are represented by the inter-coalition interference between devices.

Deviation is a key notion for the definition of the $\gamma^{\dagger}\text{-}core$. As a consequence of deviation, a new partition will be formed. Therefore, the deviation is equivalent to the formation of a new partition. In the proposed operator coalitional game,

\begin{definition} Let partition $\pi_{\mathcal{H}}$ move to $\pi^{\prime}_{\mathcal{H}}$ by deviation. 
\begin{itemize} 
\item Complete deviation: If there exists $\mathcal{D}^{C} \subseteq \mathcal{H}$ and $\pi_{\mathcal{D^{C}}} \subseteq \pi^{\prime}$ such that for all $h \in \mathcal{D^{C}}$, for all coalition $(S, \pi)$ such that $h \in \mathcal{S}, \mathcal{S} \notin \pi_{\mathcal{D^{C}}}$, then the player set $\mathcal{D^{C}}$ performs complete deviation. The players $h \in \mathcal{D^{C}}$ are called complete deviators.
\item Partial deviation: If there exists  $\mathcal{D}^{P} \subseteq \mathcal{H}$ containing only overlapping players such that for all $h \in \mathcal{D^{P}}$, for all $\mathcal{S} \in \pi_{\mathcal{H}}, \mathcal{S} \notin \pi^{\prime}_{\mathcal{H}}$, then the player set $\mathcal{D^{P}}$ performs partial deviation. The players $h \in \mathcal{D^{P}}$ are called partial deviators.
\end{itemize}
\end{definition}

In an overlapping CFG in a partition form, if a coalition of players performs a complete deviation or partial deviation, this may affect the payoffs of the remaining players. For the remaining players, we then define the \emph{residual game} as following:

\begin{definition}
Let $(\mathcal{H},v)$ be an overlapping CFG in partition form. If a subset of players $\mathcal{S}$ has already organized themselves into a certain partition $\pi_{\mathcal{S}}$. A residual game $(\mathcal{R}, v)$ is an overlapping CFG in a partition form defined on a set of players $\mathcal{R}=\mathcal{H} \setminus \mathcal{S}$. 
The players in $\mathcal{R}$ are called residuals.
\end{definition}

The residual game is an overlapping CFG in partition form on its own. In the residual game, the players react to the deviation only on the set of remaining players including partial deviators which can play further deviation.

Given two payoff vectors $\textbf{x,y} \in \mathbb{R}^{|\mathcal{H}|}$, if $\forall i \in \mathcal{S} \subset \mathcal{H}$, $x_{i} > y_{i}$, and $\exists j \in \mathcal{S}$, $x_{j}>y_{j}$, we write $\textbf{x} >_{\mathcal{S}} \textbf{y}$. Let   $(\textbf{x},\pi_{\mathcal{H}})$ denote an \emph{outcome} of the game, where $\textbf{x}$ is a utility vector resulting from a partition $\pi_{\mathcal{H}}$. 
Let $\mathbb{C}(\mathcal{H},v)$ denote the $\gamma^{\dagger}\text{-}core$ of game $(\mathcal{H},v)$, and $\Pi(\mathcal{H},v)$ denote the set of all the possible partitions of $\mathcal{H}$. 
$\gamma^{\dagger}\text{-}core$ can be found inductively in four main steps \cite{Agbaglah2009}

1) \emph{Trivial Game}: Given a coalitional game $(\mathcal{H},v)$, the  $\gamma^{\dagger}\text{-}core$ of a coalitional game with $\mathcal{H}=\{1\}$ is composed of the only outcome with the trivial partition: $\mathbb{C}(\{1\}, v)=(v(\{1\});\{1\})$.

2) \emph{Inductive Assumption}: Given the  $\gamma^{\dagger}\text{-}core$ $\mathbb{C}(\mathcal{R},v)$ for each game with $|\mathcal{R}| < |\mathcal{H}|$ players, the \emph{assumption} about the residual game $(\mathcal{R},v)$ is defined as follows: 
\begin{eqnarray}
\mathbb{A}(\mathcal{R},v)= \left\{
\begin{array}{rcl}
 \mathbb{C}(\mathcal{R},v)  ,  & &   \mathbb{C}(\mathcal{R},v) \neq \emptyset ,    \\
 \Pi(\mathcal{R},v) , & &  \text{Otherwise},
\end{array} \right. \end{eqnarray}

3) \emph{Dominance}: An outcome $(\textbf{x},\pi_{\mathcal{H}})$ is dominated via a coalition $\mathcal{S}$ if for at least one $(\textbf{y}_{\mathcal{H} \setminus \mathcal{S}}, \pi_{\mathcal{H}  \setminus \mathcal{S}} )$ there exists an outcome $((\textbf{y}_{\mathcal{S}}, \textbf{y}_{\mathcal{H}, \setminus \mathcal{S}}), \pi_{\mathcal{S}} \cup \pi_{\mathcal{H}  \setminus \mathcal{S}} ) \in \Pi(\mathcal{H},v)$ such that $(\textbf{y}_{\mathcal{S}}, \textbf{y}_{\mathcal{H}\setminus \mathcal{S} })>_{\mathcal{S}}\textbf{x}$.

4) \emph{ $\gamma^{\dagger}\text{-}core$ Generation}: The $\gamma^{\dagger}\text{-}core$ of a game of $|\mathcal{H}|$ players is the set of undominated outcomes.

The concept of dominance expresses that, given a current partition $\pi_{\mathcal{H}}$ and the respective payoff vector \textbf{x},  an undominated coalition $\mathcal{S}$ represents a deviation from $\pi_{\mathcal{H}}$ in such a way that the reached outcome $((\textbf{y}_{\mathcal{S}}, \textbf{y}_{\mathcal{H} \setminus \mathcal{S}}), \pi_{\mathcal{S} \cup \pi_{\mathcal{H} \setminus \mathcal{S}}})$ is more rewarding for the players in coalition $\mathcal{S}$, compared to $\textbf{x}$. Thus, $\gamma^{\dagger}\text{-}core$ can be seen as a set of partitions, under which the players cooperate in self-organized overlapping coalitions that provides them with the highest payoff.

During the coalition formation process of the operator coalitional game, in order to reach an outcome lies in $\gamma^{\dagger}\text{-}core$, we let each operator iteratively joins and leaves the coalitions to ensures a maximum payoff (i.e., an undominated outcome). 
To prevent loop, we introduce a variable couple \emph{history} $(H(\Gamma_{h}),H(v(\{h\})))$ for each operator to record all the coalitions that it has ever joined and the corresponding utility. If the new coalition for operator $h$ to join has been recorded and the utility that it is about to get is the same as the \emph{history}, then operator $h$ will maintain the current coalition set $\Gamma_{h}$ even if its utility will better off. Once operator $h$ changes its coalition set, the new coalition set is included in \emph{history}.
We propose the coalition formation algorithm for operators in \textbf{Algorithm 1} to reach the stable coalitional structures in $\gamma^{\dagger}\text{-}core$.

\begin{algorithm}[width=\textwidth]
\caption{Distributed Coalition Formation Game}
\label{CFG}
{\fontsize {9}{9}\selectfont
\begin{algorithmic}
\small
\STATE \textbf{Initial State} \\
\STATE \hspace{3mm} In the starting network, the operators are partitioned by $\pi_{\mathcal{H}}=\mathcal{H}={1,\cdots,H}$ with non-cooperative operators.
\STATE \textbf{Coalition Formation Process} \\
\STATE \hspace{3mm} \textbf{Phase 1} \emph{Network Discovery} \\
\STATE \hspace{6mm} Devices from the same operators perform the \emph{Dynamic Virtual 
\STATE \hspace{6mm} Link Formation} algorithm specified in  \textbf{Algorithm 2}.  Based 
\STATE \hspace{6mm} on the information feedback from device layer, each 
\STATE \hspace{6mm}  operator $h$ calculates the corresponding utility in 
\STATE \hspace{6mm} non-cooperative case.
\STATE \hspace{3mm} \textbf{Phase 2} \emph{ Coalition Formation}
\STATE \hspace{6mm} The operators play their strategies sequentially in a random 
\STATE \hspace{6mm} order.
\STATE \hspace{9mm} \textbf{repeat}
\STATE \hspace{12mm} 1) Each operators $h \in \mathcal{H}$ sequentially engages in pair-
\STATE \hspace{12mm}  wise  negotiations with another operator $h^{\prime} \in \mathcal{H} \setminus \{h\}$  
\STATE \hspace{12mm} to identify potential cooperator. During this process, 
\STATE \hspace{12mm} the devices from  the operator pair perform \emph{Dynamic}  
\STATE \hspace{12mm} \emph{Virtual Link Algorithm}. Based on the information  
\STATE \hspace{12mm} feedback, each operator calculates its potential utility. 
\STATE \hspace{12mm} 2) Based on the potential utility information and    
\STATE \hspace{12mm} \emph{history}, the pair of operators decides to form a new 
\STATE \hspace{12mm} coalition $\mathcal{S}=\{h,h^{\prime}\}$ if it ensures the utility 
\STATE \hspace{12mm} improvement.
\STATE \hspace{12mm} 3) The operators already that have cooperation with any 
\STATE \hspace{12mm} operator in $\mathcal{S}$ update their utility. Based on the 
\STATE \hspace{12mm} updated utility and \emph{history}, they perform deviation with 
\STATE \hspace{12mm} the operator(s) in $\mathcal{S}$ if it leads to utility improvement.
\STATE \hspace{9mm} \textbf{until} any further coalition formation does not result in
\STATE \hspace{9mm} utility improvement of at least one operator, i.e., conver-
\STATE \hspace{9mm} gence  to a stable partition in the  $\gamma^{\dagger}\text{-}core$.
\STATE \hspace{3mm} \textbf{Phase 3} \emph{Cooperative Sharing}
\STATE \hspace{9mm} The operators share their relay devices with cooperative   
\STATE \hspace{9mm} operators for multi-path routing according to the final 
\STATE \hspace{9mm} network graph $G^{\star}_{F}$.  
\end{algorithmic}}
\end{algorithm}

The convergence of the operator coalitional game is guaranteed due to the fact that $1)$ the total number of possible partitions with overlapping coalitions is finite, $2)$ the transition from a partition to another leads to the increase of individual utility, and $3)$ the game contains mechanism to prevent the operators to re-visit a previously formed coalitional structure. As a result, each cooperation buildup and breakup will lead to a new partition. As the number of the partition can be visited is finite, the game is guaranteed to reach a final partition, under which the utility of each operator can be no longer increased. Furthermore the last partition lies in the  $\gamma^{\dagger}\text{-}core$ because $1)$ during the coalition formation process only the partitions that bring improvement of individual utility for each operator are formed, and $2)$ in the final convergent partition, there are no dominated coalitions which the operators are better off by deviating from.

\subsection{Coalitional Graphical Game}

In the CGG, the source devices need to play \emph{transmission strategy}, while the relay devices need to play not only transmission strategy but also \emph{relay strategy}. The transmission strategy for a source device $i$ is to send a link establishment proposal to a relay $j \in \mathcal{M}^{(h)}, h \in \Gamma_{h}$ which can help to improve its payoff defined in (\ref{device}). The relay strategy for a relay device $j$ is to accept or reject a link proposal from a transmitting device $i \in \mathcal{M}^{(h)}, h \in \Gamma_{h}$ which increases its payoff defined in (\ref{Rdevice}). Once the relay device accepts the link establishment proposal, it will need to play transmission strategy like a source device. 

We consider that an device $i$ can have multiple incoming and outgoing flows simultaneously, constrained by (2), (3) and (4), and the maximum transmit power is limited to $Q^{\star}_{i}$. 
We denote $s_{i}$ the strategy space which consists of all the strategies of device $i$. When device $i$ plays strategy $s_{i}$, while all other devices maintain their current strategies denoted by a vector $s_{-i}$, the resulting network graph is denoted by $G_{s_{i},s_{-i}}$.

All the devices are considered to be myopic in the sense that each device responds to improve the payoff given the current strategies of the other devices. That is, each devices plays myopically without the foresight of the future evolution of the network. Based on this, we define the concept of \emph{best response} for devices as follows:

\begin{definition}
A strategy $s^{\star}_{i}\in \mathcal{S}_{i}$ is a best response for a device $i \in \mathcal{N}$, if $v_{\{i\}}(G_{s^{\star}, \textbf{s}_{-i}}) \geq v_{\{i\}}(G_{s_{i},\textbf{s}_{-i}})$, $\forall s_{i} \in \mathcal{S}_{i}$.
\end{definition}

Based on the concept of best response, we introduce the myopic dynamics algorithm for the proposed CGG shown in \textbf{Algorithm 2}. We define an iteration as a round of myopic plays during which each device chooses to play its current best response $s^{\star}_{i} \in \mathcal{S}_{i}$ sequentially in a random order with the aim to maximize its payoff given the current strategies of others. The dynamic virtual link formation process may consist of one or more iterations, as the best strategy of each device may change over time. All the devices play best strategies based on their flow demands. Each source device has a certain self-generated flow demand, while the flow demand of a relay device is equal to its incoming throughput. To meet the flow demand, the device on transmission may propose a link establishment proposal to multiple relay devices. The iteration stops when either all the flow demands of devices are satisfied or none of the devices can unilaterally change its strategy to improve its payoff. In other words,
when the algorithm converges, it results in a network in which none of the devices can unilaterally improve its throughput. This is referred to as a Nash network, which gives the stability concept of the final network structure $G^{\star}_{F}$. Specifically, the Nash network is defined as follows.

\begin{definition} \label{nash}
A network graph $G(\mathcal{N},\mathcal{E})$ with $\mathcal{N}$ denoting the set of all notes, i.e. devices, and $\mathcal{E}$ denoting the set of all edges, i.e, links between pairs of devices, is the Nash network in its strategy space $\mathcal{S}_{i}$, $\forall i \in\mathcal{N}$, if no device $i \in \mathcal{N}$ can improve its payoff by a unilateral change in its strategy $s_{i} \in \mathcal{S}_{i}$. 
\end{definition}
 
In the Nash network, all the links are chosen based on the best responses of devices and are thus in the Nash equilibrium. 
In a network with finite number of devices, the final network structure $G^{\star}_{F}$ resulting from the proposed CGG is a Nash network.

\begin{algorithm}[width=\textwidth]
\caption{Distributed Multi-path Routing}
\label{CGG}
{\fontsize {9}{9}\selectfont
\begin{algorithmic}
\small
\STATE \textbf{Initial State} \\
\STATE \hspace{3mm} In the starting network, each source device transmits directly to its destination device.
\STATE \textbf{Network Structure Formation Process} \\
\STATE \hspace{3mm} \textbf{repeat}
\STATE \hspace{6mm} \textbf{Phase 1} \emph{Dynamic Virtual Link Replacement}
\STATE \hspace{9mm} The devices play their strategies sequentially in a random 
\STATE \hspace{9mm} order.
\STATE \hspace{12mm} \textbf{repeat}
\STATE \hspace{15mm} 1) During each iteration $y$, every device on transmis-
\STATE \hspace{15mm} sion performs pairwise negotiation with other idle 
\STATE \hspace{15mm} devices from cooperative operators and calculates its 
\STATE \hspace{15mm} potential payoff improvement under cooperation.
\STATE \hspace{15mm} 2) After negotiation, each device $i$ plays its best 
\STATE \hspace{15mm} response $s^{\star}_{i}$, based on its flow rate demand.
\STATE \hspace{12mm} \textbf{until}  none
  of the devices can further improve its 
\STATE \hspace{12mm} payoff by a unilateral change in its strategy. 
\STATE \hspace{6mm} \textbf{Phase 2} \emph{Feedback}
\STATE \hspace{9mm} Each device $i$ sends the information about the link
\STATE \hspace{9mm} $(i,j) \in G^{\star}_{F}$ back to its operator for a coalition 
\STATE \hspace{9mm} formation decision.
\STATE \hspace{3mm} \textbf{until} Acceptance of the convergent network structure by all 
\STATE \hspace{3mm} the operators which perform \textbf{Algorithm 1}, i.e., \textbf{Algorithm 1} 
\STATE \hspace{3mm} converges.
\STATE \hspace{6mm} \textbf{Phase 3} \emph{Multi-hop Routing}
\STATE \hspace{9mm} All the source and relay devices perform multi-path 
\STATE \hspace{9mm} routing according to the final network structure $G^{\star}_{F}$.
\end{algorithmic}}
\end{algorithm}

\section{Numerical Results}
\subsection{Simulation Setting}

For simulation, we consider an LTE-A network with a TDD-OFDMA scheme, locating within a $1000m \times 1000m$ area. The bandwidth available in this network is $2MB$. The maximum transmit power of each device is $20mW$. We calculate the capacity of each link according to Shannon capacity.
The noise level is $-90dBm$. 
For the wireless propagation, as in \cite{X.Lu2014}, we set the path loss exponent $n=4$ and antenna related constant $\beta=62.5$. 
Four operators are considered in this LTE-A networks. For each operator, there is one internal flow session. For each flow session, the source device and destination device are randomly selected. The number of devices from each operator is varied from $3$ to $8$.  The results presented in this section are averaged over $1000$ times of run with random location of devices.

\subsection{Convergence Behavior of the Coalitional Graphical Game}

\begin{figure}
\centering
\includegraphics[width=0.3\textwidth]{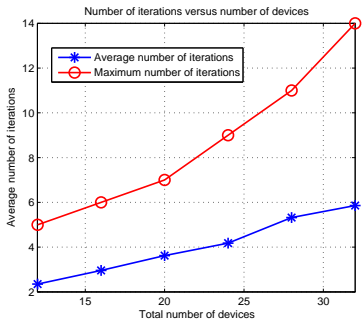}
\caption{Number of iterations versus total number of devices.} \label{average_iteration}
\end{figure}

We first examine the convergence speed of the CGG at the device layer. To this end, we set the coalition cost to be $C=0$. In this case, the grand coalition is always one of the stable coalitional structures. Hence, the simulation with this setting is equivalent to performing the CGG given the coalitional structure of grand coalition at the operator layer. Each flow is assigned with a random rate requirement within $[10, 20] kb/s$. In Fig.~\ref{average_iteration}, we show the average and the maximum number of iterations required till convergence of the algorithm from the initial network structure (i.e., direct transmission), as a function of the number of devices. 
As expected, with the increase of the number of devices, more interactions among devices is required for the CGG to converge. Observing
from Fig. \ref{average_iteration}, we find that when the total number of devices
varies from $12$ to $32$, the average and maximum numbers of
iterations vary from $2.35$ and $5$ to $5.86$ and $14$, respectively.
Thus, on average, the convergence speed of the algorithm is satisfactory.
 
 
\subsection{Case Study of Four-operator Coalition with Uniformly Distributed Devices}
We then examine the proposed LCG in a LTE-A network with uniform distribution of devices. The revenue obtained in a time unit for successfully transmitting a unit throughput (i.e., $1 Kbps$) is $120$. The operation cost $P^{C}_{i}$ is considered to be the power consumption cost. For each device to transmit or forward flow in a time unit, the cost is $500$ \emph{per} Watt.
We assume that when the operators make cooperation with another, a fixed cost $C$ is incurred to both of them. Thus, the total coalition cost afforded by operator $h$ for being in coalition $\mathcal{S}$ is $\xi^{\mathcal{S}}_{h}=C(|\mathcal{S}|-1)$. The total coalition cost for operator $h$ can be calculated by $\xi_{h}=\sum_{\mathcal{S} \in \Gamma_{h}}\xi^{\mathcal{S}}_{h}$. We set $C=5$.

\begin{figure}
\centering
\includegraphics[width=0.3\textwidth]{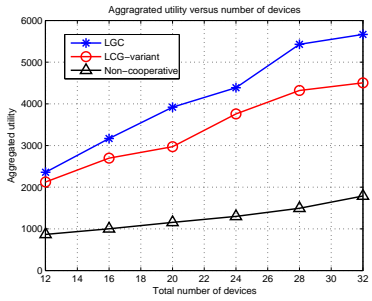}
\caption{Aggregated utility versus total number of devices.} \label{average_utility}

\end{figure}

In Fig. \ref{average_utility}, we evaluate the efficiency of the proposed LCG by presenting the average aggregated utility of all operators achieved as a function of the number of total devices. For comparison, we also study an LCG variant (labeled as ``LCG-variant") which substitutes the proposed CFG at the operator layer with a coalition formation game only enabling non-overlapping coalitions. We use the \emph{recursive core} as the stability concept for the partition of operators and adopt a solution similar to the merge-only algorithm proposed in \cite{Sudarshan} to find a stable partition. Moreover, we use the result of non-cooperative case (labeled as ``Non-cooperative"), in which the operators work independently with each individual performing CGG among its own devices, as the lower bound performance. We can observe from the figure that cooperation brings significant performance gains over non-cooperative case. The proposed LCG employing overlapping CFG outperforms the LCG variant with non-overlapping CFG. This is because the overlapping CFG allows more freedom in coalitional structure formation for potential utility improvement. In addition, we can observe that the performance gap between LCG and the LCG variant increases with the number of devices in the network.

\section{Conclusion}

The hierarchical cooperation in LTE-A networks is a promising solution for high speed data transmission and wide-area coverage. In this paper, we have presented a game-theoretic framework to model the hierarchical cooperation problem. Specifically, we have proposed a layered coalitional game (LCG) to model the cooperation behavior among the operators and devices in the different layers of LTE-A networks. The concept of extended recursive core has been advocated as the solution of stable coalitional structures. We proposed an overlapping coalition formation game for operators to find a stable coalitional structure lies in the extended recursive core that benefits all the cooperative operators. While, a coalitional graphical game has been introduced for devices to form the stable network structures for multi-path routing. 
Numerical results have shown that the proposed LCG yields notable gains relative to both the non-cooperative case and a LCG variant. The future work will characterize the performance gap between the proposed LCG and the optimal solutions obtained by centralized approaches.

\section*{Acknowledgements}
This work was supported in part by Singapore MOE Tier 1 (RG18/13 and  RG33/12),


\end{document}